\documentclass[useAMS,usenatbib]{mn2e}
\usepackage{latexsym,graphics,epsfig}
\def \be{\begin{equation}}
\def \ee{\end{equation}}
\def \bea{\begin{eqnarray}}
\def \eea{\end{eqnarray}}
\def\etal{{et al.\ }}

\title[Preheating in groups and Lithium abundance]{
Cosmic rays, lithium abundance and excess entropy in galaxy clusters
}

\author[Biman B. Nath, Piero Madau \& Joseph Silk]{Biman B. Nath$^1$, 
Piero Madau$^2$ \& Joseph Silk$^3$\\
1.  Raman Research Institute, Sadashiva Nagar, Bangalore 560080, India\\
2. Department of Astronomy and Astrophysics, University of
California, Santa Cruz, CA 95064, U.S.A.\\
3.  Department of Astrophysics, University of Oxford, Denys Wilkinson Building, Keble
Road, OX1 3RH Oxford, UK}

\begin{document}
\maketitle

\begin{abstract}
We consider the production of $^6$Li in spallation reactions by cosmic rays
in order to explain the observed abundance in halo metal-poor stars. We
show that heating of ambient gas by cosmic rays is an inevitable consequence of this 
process, and estimate the energy input required to reproduce the observed
abundance of $^6$Li/H$\sim 10^{-11}$ to be of order a few hundred eV per particle. 
We draw attention to the possibility that this could
explain the excess entropy in gas in galaxy groups and clusters. 
The evolution of $^6$Li and the accompanying heating of gas
is calculated for structures collapsing at the present epoch with injection of cosmic rays
at high redshift. We determine the energy required to explain the abundance
of $^6$Li at $z \sim 2$ corresponding to the formation epoch of halo metal-poor
stars, and also an increased entropy level of $\sim 300$ keV cm$^2$ necessary
to explain X-ray observations of clusters. 
The energy budget for this process is consistent with the expected energy output
of radio-loud AGNs, and the diffusion length scale of cosmic-ray protons
responsible for heating is comparable to the size of regions with excess entropy.
We also discuss the constraints imposed by the extragalactic gamma-ray background.

\end{abstract}
 

\section{Introduction}

The primordial abundance of lithium as predicted from the big bang 
nucleosynthesis (BBN) poses a puzzle when compared with observed abundance in
our Milky Way halo. Since lithium, along with other light elements such 
as beryllium and boron, are not produced during stellar nucleosynthesis,
the study of their abundance offers  a unique probe of the conditions
in the early universe. The agreement between recent WMAP determination
of the cosmic baryon density (Spergel  et al 2003), and the BBN 
prediction from observed deuterium and $^4$He abundances
(Cyburt et al 2005) has been encouraging, pushing one to study
other light elements. 
The predicted abundance from
BBN using the WMAP baryon density of $\Omega_b h^2=0.0224 \pm 0.0009$ is
$^7$Li/H $\approx 4.26^{+0.91} _{-0.86} \times 10^{-10}$ (Cyburt et al 2005).
A wealth of data on $^7$Li abundance among halo metal-poor stars
([Fe/H]$ \le -1.5$) has accumulated since the original discovery of a
plateau in the $^7$Li abundance by Spite \& Spite (1982).
 Current measurements find the $^7$Li abundance to be a 
factor $\sim 3\hbox{--}4$ lower
than this prediction (Asplund et al 2005; Cyburt 2004; Coc et al 2004;
Serpico et al 2004).
Cyburt et al (2005, and references therein) have recently discussed
the roles of several process including stellar destruction and new physics
to account for this discrepancy.

$^6$Li, on the other hand, is difficult to detect owing to the 
small difference in its mass from
the predominant isotope of $^7$Li, making lines from these two isotopes blend
easily. Until recently it was measured in only three stars with metallicity
[Fe/H] $\le -1.3$. Recent
high resolution studies have significantly expanded this dataset to 24 
metal-poor 
halo stars (Lambert 2004; Asplund et al. 2005), 
revealing a plateau
at $\log (^6$Li/H$)+12=0.8$ for metallicities [Fe/H] $\le -2.5$. This is
$\sim 1000$ times the predicted abundance from BBN, $^6$(Li/H)$_{\rm BBN} 
\approx 10^{-14}$ (Thomas \etal 1993; Vangioni-Flam et al 1999). 
The observed $^6$Li abundance therefore poses a challange in understanding 
its origin.  


The mismatch between the predicted and observed abundance of $^6$Li
remains puzzling even when one considers other possible production mechanisms.
$^6$Li can be synthesized after BBN epoch by
spallation (e.g., $p+{\rm O} \rightarrow$ fragments) and fusion ($\alpha + \alpha
\rightarrow ^6$Li and $^7$Li) reactions, when high-energy cosmic-ray
particles collide with ambient gas particles. Spallation reactions would
also produce B and Be in addition to lithium, and the pioneering work by
Reeves \etal (1970) matched the abundances of $^6$Li,
Be and $^{10}$B with solar system abundances.
Current models of Galactic cosmic-ray spallation predict a $^6$Li/H
abundance that is roughly proportional to the stellar metallicity (Fe/H),
with $^6$Li/H $\sim 10^{-11}$ for [Fe/H]$\sim -2$ and lower abundance
for lower metallicities ( Fields \& Olive 1999a, 1999b; 
Vangioni-Flam et al 1999;
Ramaty et al 2000; Alibes
et al 2002). The measured $^6$Li plateau in stars with
metallicities [Fe/H]$\le -2.5$ therefore cannot be explained by Galactic
cosmic-ray spallation (Lambert 2004), and requires a different origin. In
this context, Suzuki \& Inoue (2002) proposed fusion reactions involving
cosmic rays accelerated in shocks induced in the assembly of the Galactic halo,
while Rollinde et al (2005) have considered cosmological and Galactic cosmic
rays for lithium production. Prantzos (2005) has also discussed the 
energy requirements for these processes (see also Ramaty et al 2000;
Fields et al 2001).
Fields \& Prodanovi\'{c} (2005) have studied the gamma
ray background that would ensue from the same cosmic-ray population that
is expected to produce $^6$Li.
Recently Jedamjik (2004) and Kawasaki \etal (2005) have invoked
decaying relic particles to explain the $^6$Li discrepancy.
Reeves (2005) has suggested that the early massive metal-poor 
Population III stars
considered responsible for the reionisation of the universe 
injected  via their winds the low-energy cosmic rays that 
generated  $^6$Li/H via spallation in the intergalactic medium (see also
Rollinde et al 2005).

In spallation  reactions by cosmic rays, it is the low-energy $\alpha$-particles
(with kinetic energy per nucleon  $E \le 100$ MeV)
that contribute most to the production of $^6$Li as 
the cross-section decreases sharply with increasing energy (Mercer et al. 2001).
Low energy cosmic rays also deposit a large fraction of their energy into the
ambient medium through Coulomb interactions (e.g., Mannheim \& Schlickeiser 
1994). Cosmic rays as a source of heating of the intergalactic gas have been previously
studied by Ginzburg \& Ozernoi (1966) and Nath \& Biermann (1993). Moreover an 
important source of  $^6$Li is via $\alpha+\alpha$ spallation, and hence 
independent of metallicity. It is then expected that the production of 
$^6$Li from pregalactic cosmic rays could be
accompanied by substantial heating of the pregalactic  gas. 

X-ray observations of galaxy groups and clusters indeed suggest that the gas in
these objects has been pre-heated by some non-gravitational process. The
gas entropy ($S\equiv T/n_e^{2/3}$) in galaxy groups appears to be in excess 
of expectations from gravitational
interactions of gas with dark matter, and this excess entropy makes the
gas less luminous than expected (Ponman et al. 2003, and references therein).
The proposals to explain this entropy floor include energy input from
supernovae (e.g., Wu \etal 2000), warm-hot 
intergalactic medium (Valageas \etal 
2003), radiative cooling (e.g., Voit \& Bryan 2000), accretion shocks (e.g.,
Tozzi \& Nulsen 2001), and AGN heating (e.g., Roychowdhury et al 2004). 
It has been estimated that to reproduce the observed density and temperature 
profiles,
entropy must be injected into the gas at a level of $S\sim 300 $ keV cm$^2$ 
(McCarthy et al 2002).

In this paper, we assess the possibility of cosmic rays simultaneously
producing the observed abundance of $^6$Li and preheating the ambient gas
to explain X-ray observations of galaxy clusters. 
All results shown below assume a $\Lambda$CDM cosmology with $h=0.7$,
$\Omega_m=0.29$, $\Omega_\Lambda=0.71$ and $\Omega_b h^2=0.02$.

\section{Cosmic ray heating and $^6$Li abundance}

We assume that the cosmic-ray injection spectrum for protons
is a power-law in momentum, of the form
\begin{equation}
n_{\rm cr,p} dE \approx A_{\rm cr} (E+E_0) [E (E+2E_0)]^{(-\alpha +1)/2} dE \,, 
\end{equation}
where $E$ is kinetic energy per nucleon, $E_0=939$ MeV
is the nucleon rest energy, and $A_{\rm cr}$ is the normalization factor.
This form of cosmic-ray spectrum is expected from shock acceleration theories
(e.g., Blandford \& Eichler 1987). We normalize the spectrum in terms
of the energy density in cosmic-ray particles, $\epsilon_{\rm cr}$ 
(erg cm$^{-3}$)
assuming a lower limit to the kinetic energy $E_l$. 
The heating rate due to Coulomb interactions of a fully ionized
gas with cosmic-ray particles with charge $Ze$ 
is given by (Mannheim \& Schlickeiser 1994, eq. 4.22),
\be
{dE\over dt}(\beta)=4.96 \times 10^{-19}\,{\rm erg~s^{-1}}~(n_e/ {\rm cm^{-3}})
\,Z^2 \,{ \beta^2 \over \beta^3 +x_m^3},
\label{eq:loss}
\ee
where $x_m=0.0286 (T/2 \times 10^6 \, {\rm K})^{1/2}$, $T$ and $n_e$ are the ambient
(electron) temperature and density, and all other symbols have their usual meanings. 
The heating rate increases with decreasing $\beta$ but drops
below $\beta \le x_m \sim (0.5\hbox{--}3) \times 10^{-2}$ (for $T\sim 10^{4.5
\hbox{--}6}$ K). Since the gas temperatures considered below lie in the range
of $T\sim 10^{4.5
\hbox{--}6}$ K, for simplicity we assume the heating rate to be negligible
for $\beta \le 3 \times 10^{-2}$. Above this energy the rate is then independent 
of gas temperature.
The total heating rate per unit volume can be written as
\bea
\Gamma&&= \int_{E_l} n_{\rm cr,p} (E) {dE\over dt}(\beta)\, dE \nonumber\\
&& \approx  1.72 \times 10^{-15} \, {\rm erg} \, {\rm s}^{-1} \,{\rm cm^{-3}} \, 
({\epsilon_{\rm cr} \over {\rm erg} \, {\rm cm}^{-3}})
\, ({n_e \over {\rm cm}^{-3}}) \,,
\label{eq:heatex}
\eea
for $E_l=30 $ MeV, and $\alpha=2.5$.
This can be compared with the rate at fixed proton energy $E_l$
from Ginzburg \& Ozernoy (1966), 
$\Gamma \sim 6 \times 10^{-14} \, {\rm erg~s^{-1}~cm^{-3}}\,$ $(n_e/ {\rm cm^{-3}})\,
(\epsilon_{\rm cr}/ {\rm erg} \, {\rm cm}^{-3})\,$ $(E_l/30\,{\rm MeV})^{-3/2}$,
which therefore provides an upper limit to the true heating rate for a distribution
of proton energies.


The cross-section for the production of $^6$Li by $\alpha$-$\alpha$ interactions
is given by (Mercer et al 2001), $\sigma\sim 66\,\exp 
(-0.016 E_{\alpha}/ \, {\rm MeV})\,$ mb,
where $E_{\alpha}$ is the total kinetic energy of the cosmic-ray 
$\alpha$-particle. This reaction has a threshold of $\sim 10$ MeV
per nucleon.
For the abundance of $\alpha$-particles in cosmic rays, we use,
\be
n_{\rm cr,\alpha} (E) dE =  K_\alpha n_{\rm cr,p} (E) dE \,,
\ee
where, again, $E$ is the kinetic energy per nucleon, and $K_\alpha
\sim 0.08$ is the abundance factor for $\alpha$-particles. We also use
an abundance of helium in the ambient gas of $n_{\rm He}/n_{\rm H} \sim 0.08$.
The production rate of $^6$Li per proton is then written in terms of the CR energy
density as,
\bea
{d ({\rm Li/H}) \over dt} &&= \int_{E_l}  \, \sigma \, \beta c \,
K_\alpha {n_{\rm He} \over n_{\rm H}} n_{\rm cr,p} (E) 
dE  \nonumber\\ && \approx
1.05 \times 10^{-16} \, {\rm s^{-1}} 
\, ({\epsilon_{\rm cr} \over 
\, {\rm erg} \, {\rm cm}^{-3}}) \,,  
\label{eq:lithex}
\end{eqnarray}
where we have used $\alpha=2.5$ and $E_l=30$ MeV for the second equality.
Eliminating the cosmic-ray energy density from equations (\ref{eq:heatex}) and 
(\ref{eq:lithex}), we find that the production
of lithium via fusion reaction is accompanied by the deposition in the ambient
gas of the following amount of energy {\it per particle}:
\be
\Delta E \sim 16.5 \,\, {\rm erg} 
 \, \Delta ({\rm Li/H}) 
\ee
(again for $\alpha=2.5$ and $E_l=30$ MeV).
This estimate can be somewhat larger for a spectrum flatter than $\alpha=2.5$
[e.g. $\Delta E \sim 20.4$ erg $\Delta$(Li/H) 
for $\alpha=2$], or for a spectrum with a higher low-energy cutoff
[$\Delta E \sim 56.1$ erg $\Delta$(Li/H) for $E_l=50$ MeV and $\alpha=2.5$]. 
Clearly, an abundance of (Li/H)$\sim 10^{-11.2}$ through fusion reaction
should be accompanied by the deposition of $\sim 100 \, \eta $ eV per particle, where
$\eta \sim 1\hbox{--}5$ is a factor that describes the uncertainty 
in the cosmic-ray spectrum. The production of $^6$Li would not have been pervasive 
in the intergalactic medium, as cosmic rays must have been accelerated within 
(or in the vicinity of) bound structures (note that, even if this energy deposition 
had a large volume filling factor, it would occur too late
to cause a large Compton distorsion of the cosmic microwave background).
Interestingly, the above estimate of the heating associated with  
$^6$Li production is comparable to the energy needed to explain X-ray observations 
of groups and clusters, which require about $0.5\hbox{--}1$
keV per particle (Cavaliere \etal 1999; Borgani et al 2002). 
{\it Therefore the production of 
$^6$Li through spallation  reactions can have important implications in the
evolution of gas in galaxy groups and clusters.}

\section{Evolution of entropy and $^6$Li abundance}

We next calculate the evolution of gas entropy 
and $^6$Li/H abundance as a function of time. We have already mentioned X-ray data
requiring gas in groups and galaxy clusters that collapse at the present epoch 
to be endowed with an entropy $\sim 300$ keV cm$^2$. For the lithium abundance 
observed in metal-poor stars,
the constraint on the redshift of enrichment is somewhat uncertain.
The ages of halo stars are estimated to be in the range $12.5\pm3$ Gyr (e.g., 
Cowan et al 2002). In the adopted cosmology, these correspond to 
a formation redshift $\ge 1.6$. In the following, we require
that the $^6$Li/H abundance must grow to $10^{-11.2}$ by $z\sim 2$,
and that the sources of cosmic rays are confined
within structures that decouple from the Hubble flow at some point in the past
and collapse at the present epoch. We thus calculate the growth of $^6$Li/H
abundance and gas heating within such structures. In the spherical top-hat model, 
the equation of motion of a bound shell of matter is
\be
\ddot{r}=-{GM \over r^2} + {\Lambda r \over 3} \,,
\ee
where $\Lambda=3 \Omega_\Lambda  H_0^2$. Since the term with $\Lambda$
becomes smaller than the first term once the overdensity is larger than 
$\sim (2\Omega_\Lambda -1)=0.4$, we neglect
it for simplicity, and use the parametric solution $r=2r_{\rm vir} [(1-\cos
\theta )/2], t=t_c[(\theta-\sin \theta)/2 \pi]$. Here, $r_{\rm vir}=r_{\rm max}/2
={1 \over 2} [(2 G M t_c^2)/\pi ^2]^{1/3}$
is the radius at virialization, and $t_c$ is the age of the universe at
collapse, which is taken to be the present age of the universe. The density
of gas after virialization is assumed to be constant and fixed by the 
overdensity expected in the spherical top-hat model. The overdensity
for virialization in a $\Lambda$CDM universe is given by
$\Delta_c\approx 18\pi^2 + 82x- 39 x^2$ with $x \equiv 
\Omega_m(z) - 1$ (Bryan \& Norman 1998). This overdensity is $\approx 100$ 
for $z=0$. We assume the gas fraction to be $\Omega_b/\Omega_m \sim 0.14$.

For simplicity we assume the injection of cosmic rays occurs in a burst at a 
redshift $z_{\rm in}$ with an energy density $\epsilon_{\rm cr}(z_{\rm in})$.
The gas is assumed to be ionized with an initial temperature of $10^4$ K and 
an initial BBN abundance of $^6$Li/H$=10^{-14}$. The volume dilution factor
of cosmic-ray particles is calculated using the above mentioned scaling for
radius. 
We calculate the energy losses of cosmic-ray protons and $\alpha$-particles
according to equation (\ref{eq:loss}), with an additional term $d\beta 
/dt=(dz/dt)\beta (1-\beta ^2)/(1+z)$ until the turnaround epoch (when $r=r_{\rm max}$) 
to account for the losses of momentum due to the Hubble expansion. We also calculate 
the continuous depletion of $\alpha$-particles in the cosmic rays due to fusion of 
$^6$Li. We do not calculate the accompanying production of $^7$Li since the production
rates of $^6$Li and $^7$Li are similar and the additional production 
of $^7$Li/H due to cosmic rays ($\sim 10^{-11}$) is negligible compared to the
BBN value of $^7$Li/H$\sim 10^{-10}$ (Rollinde et al 2005).  

The free parameters in the calculations are the cosmic-ray spectral
index $\alpha$, the low-energy cutoff $E_l$\footnote{For simplicity 
we have assumed the same low-energy cutoff for both protons
and $\alpha$-particles, although in principle they could be different 
(Nath \& Biermann 1994).}, and the cosmic-ray
energy density at injection, $\epsilon_{\rm cr}(z_{\rm in})$. We find that 
one can reproduce the required $^6$Li/H abundance at $z\sim 2$ and the entropy 
of cluster gas at $z\sim0$ with different combinations of $\alpha, \, E_l$ and $
\epsilon_{\rm cr}(z_{\rm in})$.
Values of $\alpha$ and $z_{\rm in}$ in the range $\alpha=(2.2\hbox{--}2.7)$ 
and $z_{\rm in}\sim 3\hbox{--}6$ require $E_l=(50\hbox{--}75)$ MeV
and $\epsilon_{\rm cr}(z_{\rm in})=10^{-12.2}-10^{-11.8}$ erg cm$^{-3}$.
The requirement $E_l \sim 50\hbox{--}70$ MeV can be succintly expressed
in terms of the quantity of matter (grammage) the cosmic rays need to traverse 
before suffering large ionization losses. 
Equation (\ref{eq:loss}) can be rewritten in terms of the grammage ($\sim n m_p\beta ct$)
as $\Delta \beta/\beta \approx 6.5 \times 10^{-3} [(1-\beta^2)^{3/2}/\beta ^4] 
\times {\rm grammage}\,$, 
which shows that a $60$ MeV proton ($\beta\sim 0.341$) will lose most of its energy 
after a grammage of $\sim 2.5$ g cm$^{-2}$. For comparison, the grammage inferred
for GeV cosmic rays inside the Milky Way is $\sim 10$ g cm$^{-2}$ (e.g., Brunetti \&
Codino 2000).  We note that observed ionization rates in diffuse clouds in
our Galaxy imply $E_l \sim 30\hbox{--}60$ MeV for $\alpha\sim 2.7$
(Nath \& Biermann 1994).

\begin{figure}
\centerline{
\epsfxsize=0.5\textwidth
\epsfbox{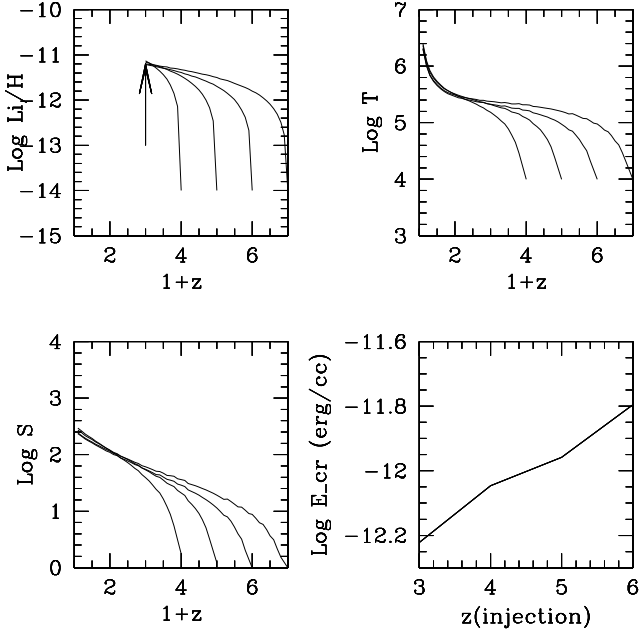}
}
{\vskip-3mm}
\caption{
The evolution of the $^6$Li/H abundance (top left panel), gas temperature (in Kelvin, top
right panel) and entropy (in keV cm$^2$, bottom left panel) with redshift 
is shown for a few cases with different $z_{\rm in}$. The values of
cosmic-ray energy density at injection,
$\epsilon_{\rm cr}(z_{\rm in})$, have been chosen to satisfy the observational
constraints of lithium abundance and gas entropy and are shown (in physical 
coordinates) in the bottom right panel.
}
\label{f:evol}
\end{figure}

\section{Discussion}

Figure 1 shows the evolution of the $^6$Li/H abundance,
gas temperature and entropy with redshift,
for various values of $z_{\rm in}$ and $\epsilon_{\rm cr}(z_{\rm in})$.
The latter have been chosen in order to produce an abundance of $^6$Li/H$\sim10^{-11.2}$ 
at $z\sim2$ and to reach $S\equiv T/n^{2/3}
\sim300$ keV cm$^{-2}$ at $z\sim0$. The required cosmic-ray energy density is
also plotted. We have used $\alpha=2.5$ and the appropriate values of 
$E_l=(50\hbox{--}75$ MeV for $z_{\rm in}=3\hbox{--}6)$.
The rate of production of $^6$Li falls rapidly after the
redshift of injection $z_{\rm in}$ because of the dilution of 
cosmic-ray energy density by the Hubble expansion, as was first noted by 
Montmerle (1977). The
abundance of $^6$Li therefore reaches a plateau soon after $z_{\rm in}$. 
Unlike the production of $^6$Li, the effect of gas heating peaks at much later epochs. 
This happens because (1) as protons lose energy in Coulomb interactions (and also 
initially due to momentum losses from the Hubble expansion), the fraction of
energy deposited into the ambient gas increases (as shown by eq. \ref{eq:loss} for 
$\beta>x_m$), and (2) $\alpha$-particles lose
energy faster (because of the $Z^2$ factor in eq. \ref{eq:loss}) and
sink below the threshold of $E\sim 10$ MeV per nucleon for fusion reactions. The 
timescale over which a cosmic-ray
proton loses all its energy is long, $t_{\rm cr} \sim 9 \times 10^{9} (E/10 \, 
{\rm MeV})^{1.5} \,
(n/10^{-5}\,{\rm cm^{-3}})^{-1} \,$ yr. 
The sharp rise in the temperature at very low redshift as
seen in Figure 1 is mostly due to adiabatic compression after turn-around 
(at $z\sim0.25$). The modest rise in entropy seen at low redshift is due to 
residual cosmic-ray protons. 

We can estimate the extent of the region heated by cosmic rays from
the distance travelled by energetic protons before ionization losses, which
depends on the diffusion coefficient. Studies of radio observations of the 
Coma cluster suggest that the diffusion coefficient of GeV particles in
a magnetic field of $B \sim 2\,\mu$G is $D \sim 
(1\hbox{--}4)\times 10^{29}$ cm$^2$ s$^{-1}$ (Schlickeiser \etal 1987).
The diffusion coefficient is proportional to the particle mean free path, which in
turn depends on the spectrum of turbulence in the ambient medium. If the
turbulence has a Kolmogorov spectrum, then the diffusion coefficient
for scattering off the magnetic irregularities
scales as $D \propto r_g^{2-5/3} \propto (E+E_0)^{1/3} B^{-1/3}$, 
where $r_g$ is the
gyration radius, $E$ and $E_0$ are kinetic and rest energy of 
particles (Biermann \& Strittmatter 1987; Berezinsky \etal 1997). For cosmic-ray 
protons responsible for heating, $E\sim E_0$.
We can estimate the diffusion coefficient of these protons
for $B \sim 10^{-9}$G  
as $D \sim 2.5 \times 10^{30} (B/10^{-9} \, G)^{-1/3}$
 cm$^2$ s$^{-1}$. The distance travelled in $t \sim 10^{10}$ yr is
then $r \sim \sqrt{(6 D t)} \sim 0.7 (B/10^{-9} \, G)^{-1/6} \, 
(t/10^{10} \, {\rm yr})^{1/2}$ Mpc. Although
the scalings with magnetic field and particle energy are somewhat uncertain, we
can compare this length scale with the scale of structures considered
in this paper. 

X-ray observations of galaxy clusters indicate that gas entropy is enhanced in
the central region, with $r \sim 0.1 r_{200}$, where $r_{200}$ 
($\sim 0.9\hbox{--}1.9$ Mpc for 
a cluster of mass $M\sim 10^{14\hbox{--}15}$ M$_{\odot}$) is 
the radius within which
the mean overdensity is 200 times the ambient density, although there may be
some evidence for enhanced entropy out to 
$r_{500}$ ($ \sim 0.5\hbox{--}1.2$ Mpc for $M 
\sim 10^{14\hbox{--}15}$ M$_{\odot}$, Ponman et al 2002). We therefore
conclude that cosmic rays can heat gas in these regions in $ \sim$ a few Gyr. 

The required cosmic-ray energy density in this scenario is (from Figure 1)
is $\epsilon_{\rm cr} \sim 6 \times 10^{-13}$ erg cm$^{-3}$ at $z\sim 3$, 
so that the comoving
energy density is $\sim 9.4 \times 10^{-15}$ erg cm$^{-3}$. 
If the cosmic rays fill structures
corresponding to groups and clusters, with a volume filling factor 
$f\sim 10^{-3}$ (for structures with $M \ge 10^{14}$ M$_{\odot}$ in a 
$\Lambda$CDM universe), then the comoving
energy density is $\sim 9.4 \times 10^{-18} (f/10^{-3})$ erg cm$^{-3}$. 
If the efficiency
of accelerating cosmic rays is $\eta \sim 0.15$ then this requires an energy
density of $6.3 \times 10^{-17}(\eta/0.15)^{-1}(f/10^{-3})$ erg cm$^{-3}$. 

According to Chokshi \& Turner (1992) and Yu \& Tremaine (2002), 
the integrated energy density in radiation from quasars is
$\sim1.3 \times  10^{-15}$ erg cm$^{-3}$ at $z=0$; for quasars at $z\ge2$, 
the energy density
is $\sim 6.5 \times 10^{-16}$ 
erg cm$^{-3}$. Radio loud AGNs possibly deposit comparable mechanical 
energy in outflows as in
radiation (Furlanetto \& Loeb 2001), but only a fraction $\sim 0.1$ of AGNs are
radio-loud. 
The available energy density in outflows from radio galaxies $z\ge2$ is
therefore estimated to be
$6.5 \times 10^{-17}$ erg cm$^{-3}$, comparable to the requirement in the 
present scenario. We then find that the energy budget in this scenario 
is consistent with the energy available from
radio-loud AGNs, and the diffusion length scale of cosmic-ray protons
responsible for heating is comparable to the size of the  region with enhanced
entropy in clusters. 
Cosmic-ray particles would also be accelerated by supernovae responsible for the
enrichment of the intracluster medium. Berezinsky \etal (1997)
estmated the cosmic-ray energy in a cluster of total mass $M\sim2 \times
10^{14}$M$_{\odot}$
to be $\sim 4.4 \times 10^{60}$ erg, which corresponds to an energy density
$\sim 3.5 \times 10^{-14}$ erg cm$^{-3}$ for a region of size $\sim 1$ Mpc. This
is smaller than the energy density $\sim 5 \times 10^{-13}$ erg
cm$^{-3}$ required in our scenario, but if produced within a smaller region,
supernovae also could be an important source of cosmic
rays if produced early ($z \ge 2$).

Recent observations  (Croston et al 2004) have shown that although gas in
groups with currently active radio galaxies has high entropy, even groups 
with radio-quiet AGNs deviate somewhat from the scaling relation between the X-ray
luminosity and temperature extrapolated from rich clusters. It is conceivable
that past activity of radio galaxies could have raised the entropy in these
systems (Roychowdhury et al 2005). This observation is consistent with
the present model since the AGNs responsible for gas heating may have been 
active in the past, at $z \ge 2$, and be inactive
 today. The synchrotron life time of cosmic-ray
electrons with $\sim 10$ GeV electrons, which can radiate at $\sim 600$ MHz for $B
\sim 1 \, \mu$G, is $\sim 1.4 (E/10 \, {\rm GeV})^{-1} (B/ 1 \, \mu {\rm G})^{-2}$ 
Gyr: 
this emission might be detectable in future continuum observations at low 
radio frequencies.

In the context of the lithium abundance in the halo stars
of our Galaxy,
our scenario would require a pregalactic generation of
cosmic ray sources that might plausibly be associated with intermediate
mass black holes or miniquasars. These objects are likely precursors to
the supermassive black holes observed at present in our Galaxy and
in M31.

The cosmic-ray protons responsible for gas heating
would also interact with ambient protons to produce
pions which would eventually decay to produce gamma-ray photons. It is therefore
important to study the consequences of our scenario for the  production
of  high-energy photons.
We first estimate the emissivity of gamma radiation above a certain photon
energy $\epsilon_\gamma$ using the analytical fit to the differential 
emissivity provided by Pfrommer and En\ss lin (2004).
For cosmic rays with a power-law index $\alpha=2.5$, we estimate
the gamma-ray emissivity in the ambient gas to be (in the units of
photons s$^{-1}$ proton$^{-1}$)
\begin{equation}
f(>\epsilon_\gamma) \approx 1.2 \times 10^{-25} \Bigl ( 1+
{\epsilon_\gamma \over 0.2 
\, {\rm GeV}} \Bigr )^{-1.45} \, 
\Bigl
( { \epsilon_{\rm cr} \over
10^{-12} \, {\rm erg} \, {\rm cm}^{-3}} \Bigr ) \,.
\end{equation}
The photon number density $n_{\gamma,{\rm Li}}$ expected to accompany the fusion
of $^6$Li can then be written as,
\begin{equation}
{n_{\gamma,{\rm Li}} (> \epsilon_\gamma) \over n_p} \approx 1.16 \times 10^{-8} 
\Bigl ( 1+{\epsilon_\gamma \over 0.2 
\, {\rm GeV}} \Bigr )^{-1.45} \, \Bigl
( { \rm Li/H \over
10^{-11}} \Bigr ) \,,
\end{equation}
where we have used equation (\ref{eq:lithex}) to eliminate the dependence
on cosmic-ray energy density and time.

Prodanovi\'{c} and Fields (2004, eq. 3) provide a convenient fit to the
observed extragalactic gamma-ray background. We integrate it
to estimate the flux above $\epsilon_\gamma$ to be
$F_\gamma (>\epsilon_\gamma) \sim 5 \times 10^{-6} (\epsilon_\gamma /0.2 \,
{\rm GeV})^{-1.2}$ photons cm$^{-2}$ s$^{-1}$ sr$^{-1}$, 
for $0.05$ GeV $\le \epsilon_\gamma \le 1$ GeV. Following
Silk and Schramm (1992), and using the present-day mean proton density
$n_{p,0}\sim 1.9 \times 10^{-7}\,$cm$^{-3}$, we can write the observed gamma-ray flux as,
\begin{equation}
{n_{\rm obs} (> \epsilon_{\gamma}) \over n_{p,0}} \approx 1.1 \times 10^{-8}
(\epsilon_\gamma /0.2 \,
{\rm GeV})^{-1.2} \,.
\end{equation}
We then find that a fraction $\sim 0.15-0.2$ of the extragalactic gamma
ray background between $0.1\hbox{--}1$ GeV could have been produced by
cosmic rays responsible for the cosmological production of $^6$Li. 


\section{Summary}

We have drawn attention to the connection between the possible origin
of $^6$Li in halo metal-poor stars and the excess entropy in galaxy clusters
as indicated by X-ray observations. Spallation reactions and heat losses by
cosmic rays can reproduce the observed abundance of $^6$Li/H$\sim 10^{-11.2}$ 
by $z \sim 2$ and raise the entropy of cluster gas to $\ge 300$ keV cm$^2$ by $z\sim0$. 
The required energy budget is consistent with that expected from radio-loud AGNs.
The size of the region heated by cosmic-ray protons ($\sim 0.7$ Mpc) 
is comparable to the observed scales of gas
with excess entropy. The expected gamma-ray flux
from pion decay is a fraction $0.15\hbox{--}0.2$ of the observed extragalactic 
background at $0.1\hbox{--}1$ GeV. The scenario described in this paper then
ties together two apparently disparate sets of observations -- of enhanced
$^6$Li abundance in metal-poor stars and of enhanced entropy in gas in galaxy
groups -- with a population of cosmic rays that may have originated from
AGN activity at early times.

\bigskip

We thank Brian Fields, Peter Biermann and David Lambert 
for useful discussions and the referee for helpful comments. We are
grateful to the Kavli Institute for Theoretical Physics for hospitality 
during the beginning of this project. Support for this work was provided by NASA grants
NAG5-11513 and NNG04GK85G, and by NSF grants AST-0205738 (P.M.).

\end{document}